\documentclass[12pt,preprint]{aastex}
\usepackage{epsfig, array, amsmath, delarray}

\shorttitle{Relativistic stellar aberration..}
\shortauthors{Turyshev}

\begin{document}

\title{Relativistic stellar aberration 
 for the \\ Space Interferometry Mission: parallax analysis}

\author{Slava G. Turyshev\altaffilmark{1}}
\affil{Jet Propulsion Laboratory, California Institute
of  Technology, Pasadena, CA 91109}

\begin{abstract} 
This paper is a continuation of our previous analysis
(i.e. Turyshev 2002a, 2002b)  of the relativistic stellar aberration
requirements for the Space Interferometry Mission (SIM).   Here we have
considered a problem of how the expected astrometric accuracy of  parallax determination will constrain the accuracy of the spacecraft navigation.  We
show that   effect of the spacecraft's navigational errors on the accuracy of
parallax determination with SIM will be negligible. We discuss the implication of
the results obtained for the future mission analysis.
\end{abstract} 

 
\section{Introduction}

It is expected that SIM will achieve the $\sigma_\pi=4~\mu$as mission accuracy
in determination of parallaxes to a nearby stars. 
\cite{TuryshevUnwin98} have demonstrated  that future astrometric model for SIM
will have to account not only for the effects introduced by the motion of the
solar system bodies, but also for those that are generated by the motion of the
spacecraft  and corresponding errors in the spacecraft's navigation. 
Here we would like to analyze whether or not the future SIM astrometric accuracy
will be compromized by the spacecraft navigational errors.  

\begin{figure}[ht]
 \begin{center} 
    \psfig{figure=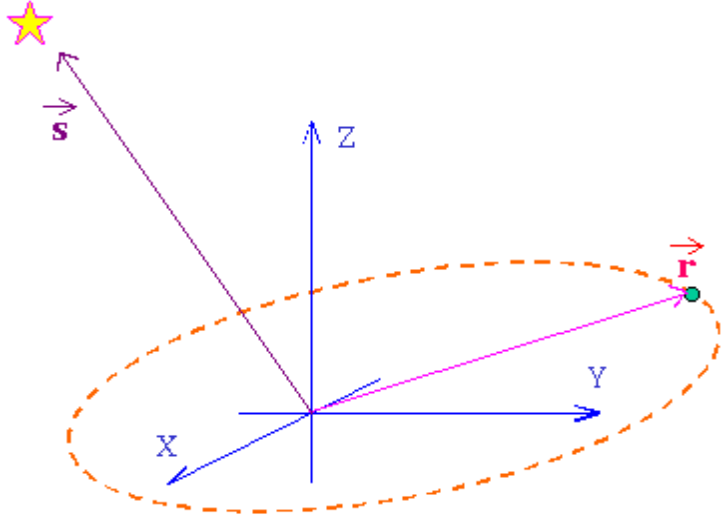,width=100mm,height=70mm}
     \caption{Geometry of the problem. 
      \label{fig:parallax}}
 \end{center}
\end{figure}

Analysis of tolerable errors for  position determination  in a
general case is a complicated problem and should be addressed by using a
formal numerical treatment. 
However, we may simplify the task by analyzing a special case which is expected
to provide the most stringent requirements on the positional accuracy. 
The three-dimensional vector of parallax, $\vec{\pi}$, is given by the
expression:
\begin{equation}
\vec{\pi}=\frac{\vec{s}\times(\vec{r}\times\vec{s})}{\cal D},
\label{eq:par0}
\end{equation}
\noindent where $\vec{s}$ and ${\cal D}$ is the direction and  distance 
to the  observed source  correspondingly; and $\vec{r}$ is the barycentric
position of the observer. The overall three-dimensional geometry of the
problem   presented in  Figure \ref{fig:parallax}. The  involved
vectors  may be given in the  spherical coordinate
system  by their  magnitudes and  two corresponding   sky-angles:
\begin{eqnarray}
\vec{s}&=&~\,\big(\cos\alpha\cos\delta,~\sin\alpha\cos\delta,
~\sin\delta\big),\nonumber\\[1mm]
\vec{r}&=&r\big(\cos\lambda\cos\chi,~\sin\lambda\cos\chi,
~\sin\chi\,\big). 
\label{eq:vect}
\end{eqnarray}

This parameterization allows one to present the expression 
Eq.(\ref{eq:par0}) in the following form:
 \begin{equation}
\vec{\pi}=\frac{r}{\cal D}~\vec{\cal P},
\label{eq:par} 
\end{equation}
\noindent with vector $\vec{\cal P}$ given by

\begin{eqnarray}
\vec{\cal P}&=&
\vec{e}_x\Bigg(\sin\alpha\,\cos^2\delta\,\cos\chi\,\sin(\alpha-\lambda)-
\sin\delta\Big[\cos\alpha\,\cos\delta\,\sin\chi-
\sin\delta\,\cos\lambda\,\cos\chi\Big]\Bigg)-\nonumber\\[1mm]
&-&\vec{e}_y\Bigg(\cos\alpha\,\cos^2\delta\,\cos\chi\,\sin(\alpha-\lambda)+
\sin\delta\Big[\sin\alpha\,\cos\delta\,\sin\chi-
\sin\delta\,\sin\lambda\,\cos\chi\Big]\Bigg)+\nonumber\\[1mm]
&+&\vec{e}_z\Big(\cos^2\delta\,\sin\chi-
\sin\delta\,\cos\delta\,\cos\chi\,\cos(\alpha-\lambda)\Big) .
\label{eq:par_par} 
\end{eqnarray}

\section{Covariance matrix}

The expressions Eqs.(\ref{eq:par})-(\ref{eq:par_par}) may now be used to answer
the question - how the positional errors   
$\sigma_r,\,\sigma_\lambda,\,\sigma_\chi$  will affect the accuracy of  
parallax determination, $\sigma_\pi$. To do this one needs to 
take the first order variations of the right-hand side of the expression
(\ref{eq:par}) with respect to  positional errors $\delta r, \, \delta\lambda$
and $\delta
\chi$. The goal here is  to present the obtained result in the form:
{}
\begin{equation}
(\delta\vec{\pi})^2 =\left(\frac{\delta r}{\cal D}\right)^2
\vec{\cal P}^2+
\frac{r^2}{{\cal D}^2}(\delta \vec{\cal P})^2+\frac{2\delta r}{\cal D}
\frac{r}{\cal D}(\vec{\cal P}\cdot\delta\vec{\cal P}).
\label{eq:var}
\end{equation}
\noindent  This expression is quite difficult for analytical description in a
general case,  however, for the purposes of  present study,  it may be 
significantly simplified. To do this, we  assume that motion of the 
spacecraft is almost in the plane of ecliptic, thus
$\chi\sim0$, but $\delta\chi\not=0$. We will not go into the lengthy
derivations, but will present here only the final results for this algebraic
exercise: {}
\begin{eqnarray}\nonumber
\vec{\cal P}^2&=&1-  \cos^2\delta\,\cos^2(\alpha-\lambda),\\\nonumber
(\delta \vec{\cal P})^2&=&
\delta\lambda^2\,\Big[1-\cos^2\delta\,\sin^2(\alpha-\lambda)\Big]+
\delta\chi^2\,\cos^2\delta - \delta\chi\,\delta\lambda\,\sin2\delta\,
\sin(\alpha-\lambda)\\ 
(\vec{\cal P}\cdot\delta\vec{\cal P})&=&
\delta\lambda\,\cos^2\delta\,\sin(\alpha-\lambda)\,\cos(\alpha-\lambda)-
\delta\chi\,\sin^3\delta\,\cos\delta\,\cos(\alpha-\lambda).
\end{eqnarray}
 
Assuming that the errors  $\delta r,\, \delta \lambda$ and $\delta\chi$ 
are normally distributed and uncorrelated,  one may average the  obtained
expressions over the period of the spacecraft's barycentric orbit. To further
simplify the analysis we will assume that the orbit is circular. The latitude
argument for circular motion coincides with the mean anomaly, e.q.
$\lambda=n\,t$, where
$n=\frac{2\pi}{T}$  and $T$ is the period of orbital motion of the
spacecraft.  As a result, one obtains the expression for contribution of the
navigational errors  to the accuracy of  parallax determination:
{} 
\begin{equation}
\sigma_\pi^2 = \frac{\sigma^2_r}{{\cal D}^2}\,\frac{1}{2}(1+\sin^2\delta)+
\frac{r^2}{{\cal
D}^2}\Big[ \,\frac{\sigma_\lambda^2}{2}(1+\sin^2\delta)+
\sigma_\chi^2 \cos^2\delta\,\Big].
\label{eq:par00}
\end{equation}
 
\section{Obtained results and conclusions}

The obtained expression represents the error in parallax determination 
depending on the accuracy of SIM navigation. We have considered the motion
of the spacecraft almost in the plane of ecliptic. The three terms on the
right-hand side of the expression (\ref{eq:par00}) represent the contributions 
of the radial, $\sigma_r$ and tangential, $\sigma_\lambda,
\sigma_\chi$, positional errors. The obtained expression valid for any source,
but depends only on $\delta$ - the declination angle of the source's position.
One may see that   positional errors are most influencing  determination
of parallax for the case when the source is near the poles, or 
$\delta\sim\pm\frac{\pi}{2}$. Thus, leading us to the  final expression:
{}
\begin{equation}
\sigma_\pi^2 = \frac{\sigma^2_r}{{\cal D}^2} +
\frac{r^2\,\sigma_\lambda^2}{{\cal D}^2}.
\label{eq:par!}
\end{equation}

The expression Eq.(\ref{eq:par!}) is somewhat familiar and it may be obtained by
a much simpler way. However, the second term on the right-hand side  of it is
very important. This term allows one to quickly obtain the necessary result.
Thus, one may show that the tangential error, $\sigma_\lambda$, is related to the
radial one as follows $\sigma_\lambda=\sigma_r/r$. One may also expect that  this
longitudinal error in the spacecraft's position  
will be the same as the error in the spacecraft's velocity  sky-angle 
$\sigma_\psi$=$\sigma_\lambda$. \cite{TuryshevUnwin98} have shwon that this error
was related to the velocity error by $\sigma_\psi=\sigma_v/v$. This is why we may
use the following relation  $\sigma_r/r = \sigma_v/v$ to obtain the  maximum
error in the SIM's radial distance from the Sun for the entire mission. 
As a result, this relation suggests that the barycentric distance of the
spacecraft during the whole mission  should be   known with an accuracy equal or
better then $\sigma_r =  20$ km.

Finally, one may obtain the  expression for the expected parallax
errors induced by the inaccuracies in determination of the SIM's solar
system's barycentric position. For example, for a source located at the
distance
${\cal D} = k$ pc ~from the Sun, the corresponding expression reads:
{}
\begin{equation}
\sigma_\pi= \frac{r}{\cal D} \frac{\sigma_v}{v}=
\frac{1.5\times 10^{13}~
{\rm  cm}}{3\times10^{18}~{\rm  cm}}\,\frac{4~{\rm  mm/s}}{3\times10^7~{\rm  mm/s}}
\,\frac{10^{12}}{4.85}=\frac{0.1}{k}~~~\mu{\rm  as}.
\label{eq:par!2}
\end{equation}

Thus, provided that the accuracy of the spacecraft's velocity determination 
(driven by the need to correct for the relativistic stellar aberration) will be
at the level of $\sigma_v=4$ mm/s, the influence of the spacecraft's barycentric
position errors    will be  negligible for the parallax determination.

\acknowledgments
The reported research   has been done at the Jet Propulsion
Laboratory,  California Institute of Technology, which is under  contract to the 
National Aeronautic and Space Administration.


\begin{thebibliography}{}
\bibitem{Turyshev021} Slava G. Turyshev:  2002a,  
     {\it Relativistic Stellar Aberration Requirements   for the
     Space Interferometry Mission}.
      Submitted for publication. May 2002. gr-qc/0205061

\bibitem{Turyshev022} Slava G. Turyshev:  2002b.  
     {\it Relativistic Stellar Aberration Requirements   for the
     Space Interferometry Mission (2)}.
      Submitted for publication. May 2002. gr-qc/0205062

\bibitem[Turyshev and Unwin (1998)]{TuryshevUnwin98}
Turyshev, S. G.  and  Unwin, S. C., 1998, 
\textit{Relativistic Stellar Aberration  Requirements for the Space
Interferometry Mission}, JPL Interoffice Memorandum: 1998-1117 (November 17,
1998) 
\end{thebibliography}
\end{document}